# The Twist Bend Nematic: A Case of Mistaken Identity.

Edward T. Samulski[a]*, Alexandros G. Vanakaras[b], and Demetri J. Photinos[b]

[a] Department of Chemistry, University of North Carolina, USA

[b] Department of Materials Science, University of Patras, Greece

email: et@unc.edu

## Abstract

We review the physics underlying Meyer's conjecture of how macroscopic-scale twist and bend conspire within the Frank-Oseen elasticity theory of nematics to create a heliconical arrangement of the uniaxial, apolar nematic director, the so-called "twist bend nematic" $N_{TB}$. We show that since 2011 a second, lower-temperature nematic phase observed in odd methylene-linked cyanobiphenyl dimers discovered by Toriumi and called $N_X$, has been incorrectly identified as $N_{TB}$. Moreover, as more quantitative data on the $N_X$ emerged, Meyer's simple prediction has been distorted to accommodate those findings. In fact, the molecular organization in the $N_X$ conforms to the $N_{PT}$ phase, a polar, twisted arrangement of nonlinear mesogens advanced in 2016. The attributes of the $N_{PT}$ are summarized and differentiated from those of the $N_{TB}$ in an effort to contribute to a better understanding of the $N_X$ phase and, equally important, to encourage researchers to continue to search for a liquid crystal that exhibits Meyer's pioneering theoretical suggestion, namely that form-chirality can exist in simple nematics composed of <u>achiral</u> molecules.

Keywords: twist bend nematic, polar twisted nematic.

At the Les Houches Summer School in Theoretical Physics in 1973, Robert Meyer discussed "static problems in liquid crystals, especially problems of structure, from the molecular to the macroscopic level." His notes, *Structural Problems in Liquid Crystals*, were subsequently published in 1976.[1] Therein he considered spontaneous polarization in the context of flexoelectricity:

> "2  *Spontaneous polarization*
> Now, assume for simplicity that **E** = 0, and that the molecular polarization does not involve electrostatic effects that would produce macroscopic electric fields. However, assume that there is a local intermolecular interaction that tends to produce a finite polarization **P**⁰. This effect can be included in the free energy as follows:

$$g = \frac{1}{2}\left(K_{11}^0 - \frac{a_1^2}{\alpha_1}\right)S^2 + \tfrac{1}{2}K_{22}t^2 + \frac{1}{2}\left(K_{33}^0 - \frac{a_3^2}{\alpha_3}\right)B^2$$
$$-a_1 P_\parallel^0 S - a_3 \mathbf{P}_\perp^0 \cdot \mathbf{B} \quad .$$
$$P_\parallel = P_\parallel^0 + \frac{a_1}{\alpha_1}S, \quad \mathbf{P}_\perp = \mathbf{P}_\perp^0 + \frac{a_3}{\alpha_3}\mathbf{B}$$

The linear terms in the free energy density, proportional to [splay] **S** and [bend] **B** indicate that the ground state for the polar nematic should now contain finite splay or bend or both, depending on the nature of the polarization. The effect is similar in principle to that found for chiral asymmetry in the case of a nematic phase. However, the geometry is quite different.

Although a state of uniform torsion is possible, a state of constant splay is not possible in a continuous three dimensional object. A state of pure constant bend is also not possible, although <u>a state of finite torsion and bend is possible</u>. The latter is a modified helix in which <u>the [uniaxial nematic] director has a component parallel to the helix axis</u>. In laboratory coordinates,

$$n_z = \cos\varphi, \quad n_x = \sin\varphi\cos t_0 z, \quad n_y = \sin\varphi\sin t_0 z.$$

The magnitude of the bend is $t_0 \sin\varphi\cos\varphi$ [where $t_0$ is the helical wavenumber].

No helical structure has ever been reported in a non-chiral nematic [as of 1973]. If it occurred in a chiral nematic, it might be difficult to distinguish from the ordinary [cholesteric] helix, without special optical equipment."

(ref 1, pp 319-320; <u>emphasis</u> added)

Robert Meyer's 1973 conjecture of a "twist-bend" helix—the spontaneous formation of a heliconical trajectory of the *apolar* nematic director **n**, the so-called twist bend nematic $N_{TB}$ (Figure 1)—lay dormant for a quarter of a century. Then in 2001 Memmer presented images [2] from Monte Carlo simulations of idealized bent-core mesogens (linked Gay-Berne particles with $C_{2v}$ symmetry) using periodic boundary conditions that defined the pitch of the helical arrangements of banana-shaped mesogens. Those images (e.g., ref 3, Fig. 11), published in 2002,[3] appear to have reinforced Dozov's use of a simple Landau-like phenomenological model of an apolar uniaxial nematic to show symmetry-breaking transitions in the nematic phase.[4] But Memmer's pitch scale was determined *a priori* by merely setting it equal to the simulation box length; he finds $p \sim 40$ molecular lengths, which is an artifact of the number of molecules that fit into his simulation box. [5] When discussing helical superstructures he alludes to chiral "domains with a so-called *twist-bend structure* … [with] the local director **n** spiraling around the helical axis …with constant tilt angle and pitch," citing analogies with the layered chiral smectic C* phase and a theoretical model for cholesterics with conic supramolecular organization described by Pleiner and Brand.[6] Whereas Dozov computes a helical pitch $p \sim 300$ nm, categorized as "rather small but still macroscopic," it is somewhat smaller than the pitch implied by Meyer for the $N_{TB}$ (see Fig. 1), but clearly one or more orders of magnitude larger than the molecular dimensions of the constituent nematogens. Despite the pitch scale discrepancies, those two "reinventions" [7] of Meyer's $N_{TB}$ model predisposed the liquid crystal research community to consider experimental observations of twisted supramolecular organization in nematic phases comprised of achiral nematogens as potential evidence for the $N_{TB}$ phase. The ultimate target of this predisposition was the second, lower-temperature nematic phase in methylene-linked cyanobiphenyl dimers (CB-C$n$-CB; for $n$ odd) first discovered in 1991 by Toriumi.[8] This dimer system was subjected to intense scrutiny in general because of its second nematic phase but, more specifically, because there was an ongoing search for a macroscopic biaxial nematic. The prevailing thinking at the time was that biaxial nematic phases were not observed because on cooling, candidate nematics were intercepted by smectic phases and still considered to be an experimental limitation, one that can be obviated in simulations. [9] And while the lower temperature nematic in Toriumi's odd-dimer system was initially reported to be smectic,[10] a more thorough study showed that the low temperature phase was nematic thereby increasing interest in this second nematic phase. [11] That low temperature nematic was designated $N_X$. Subsequent NMR observations [12,13] indicated that the phase could discriminate among enantiotopic deuterons, i.e., NMR showed that some sort of chiral supramolecular arrangement is present in the Nx phase. In 2011, ref [12] equated that apparent chiral structure in the $N_X$ phase to Meyer's $N_{TB}$ phase. Notably, no indication or reference as to the possible length-scale of the helical pitch is made in this, otherwise very extensive, work. That (erroneous) association—$N_{TB}$ = $N_X$—launched an international rush to publish about a variety of observations intended to corroborate the putative discovery of the twist bend nematic, one that continues in 2020.

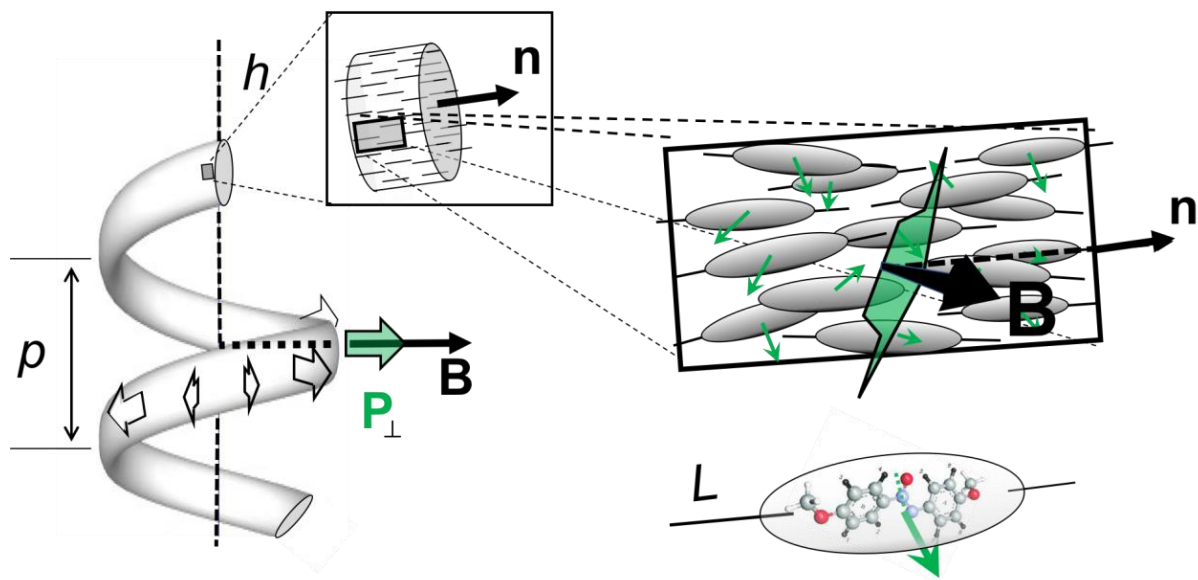

**Figure 1. Schematic diagram of the twist bend nematic.** Using a typical prolate, lath-like calamitic nematogen with transverse electrostatic polarity and long axis *L*, we schematically indicate the helical trajectory of the director **n** about a helix axis *h* with a pitch $p \gg 1$ $\mu$m and local (flexoelectric) polarization $\mathbf{P}_\perp$, remaining normal to *h* (block arrows) and coinciding with the direction of the respective bend vector $\mathbf{B} = \mathbf{n} \times \nabla \times \mathbf{n}$.

Despite the irreconcilable differences between the length scales inherent in Meyer's conjecture—applicability of the continuum Frank-Oseen elasticity theory—and the pitch exhibited by the $N_X$ phase of CB-Cn-CB dimers ($p < 10$ nm),[14] some researchers continue to identify the $N_X$ phase as the $N_{TB}$ phase. Such an obvious and fundamental incompatibility failed to arrest proposals for the local structure in the $N_x$ phase. Instead there was a concerted attempt to force-fit observations/modeling to Meyer's proposal and/or subsequent reinventions thereof. Here, in an effort to attenuate the propagation of the misnomer "twist bend nematic" for the $N_X$ phase—there are journal issues devoted to this misidentified phase, [15] conference reports, [16] high-profile perspectives, [17] the latest edition of the Handbook of LCs [18], and recent reviews [19]—we point to compelling evidence that the $N_X$ phase does not conform to Meyer's twist bend conjecture.

First, the length scale of the $N_X$ pitch precludes the applicability of the Frank-Oseen elasticity on which the $N_{TB}$ model is based (either using the flexoelectric formulation[1] or based on the negative value of the bend elastic constant[4]). Secondly, the enantiotopic discrimination data, which initiated the erroneous assignment of the $N_X$ as a $N_{TB}$, was shown not to constitute proof of any heliconical structure of the nematic director[13]. Thirdly, it was later demonstrated that the enantiotopic discrimination exhibited by small rigid solutes in the $N_X$ can be consistently accounted for by the combination of polar and tilted local ordering of the molecules[20] and that no enantiotopic discrimination can be accounted for by the $N_{TB}$ model. Fourthly, a molecular simulation of the CB-n-CB dimers[21] showed a lower temperature, positionally disordered, phase of short-pitch (<10nm) modulated ordering for the odd-n members, but without any local symmetry axis that meets the requirements of a nematic director **n**; on the contrary the ordering shows strong polarity, with the polar direction tightly twisted at right angles to a well-defined helical axis.

If not an $N_{TB}$ phase, what is the nature of the lower temperature nematic phase, Nx, exhibited by the odd homologues of the methylene-linked cyanobiphenyl dimers? All of the key attributes of the $N_X$ phase are readily accounted for by a new nematic phase model, the polar twisted nematic ($N_{PT}$) advanced by Vanakaras and Photinos, refs [21,22]. Their polar twisted nematic $N_{PT}$ has a supramolecular arrangement wherein the coarsely V-shaped dimer molecules exhibit polar ordering along a direction that is *tightly* twisted at right angles to a macroscopic helical axis *h* (see Figure 2). According to this polar-twisted nematic model, the direction of polar molecular ordering—the local director **m**, a two-fold symmetry axis of the phase and the only director (see Figure 2 inset)—tightly twists about *h*, the "helix axis", that in turn, is the only macroscopic axis of full rotational symmetry.

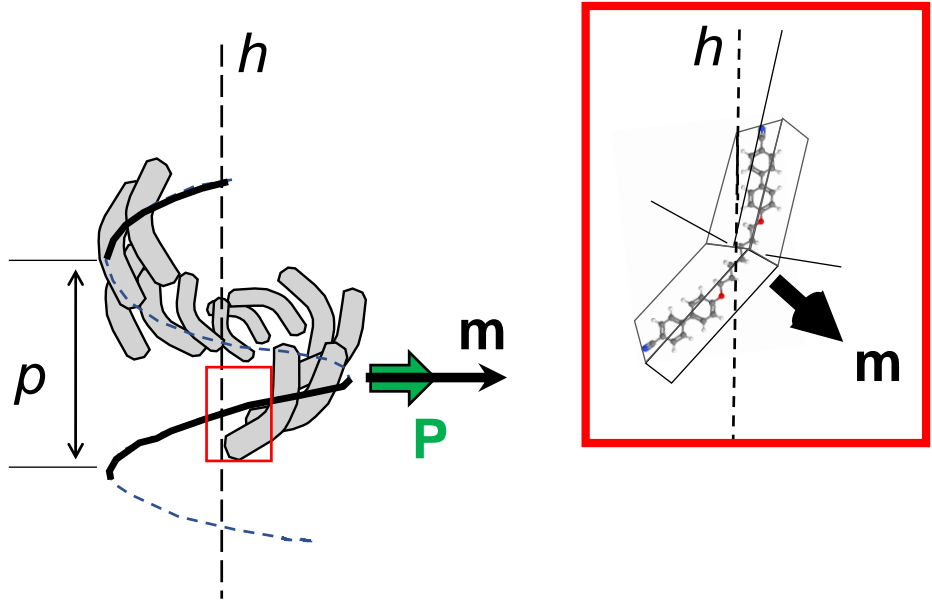

**Figure 2. Schematic polar twisted phase adopted by CB-Cn-CB dimer mesogens.** In the $N_{PT}$ phase the pitch $p \sim 10$ nm. The polarization **P** is coincident with **m** and normal to $h$; **P** arises because of the local polar ordering of the steric/electrostatic molecular profile of V-shaped mesogens. The polar director **m**, which is a local $C_2$ symmetry axis of the phase, undergoes pure twisting about $h$. In the idealized dimer (insert), its steric/electrostatic polar axis aligns preferentially along **m** with its (statistical) plane of symmetry tilted relative to $h$ (red shading).

Obviously, the presence of a polar director in the $N_{TB}$ would be in direct contradiction with the assumed full rotational symmetry about the local nematic director **n** (and its apolarity, in the sense of the equivalence between **n** and **−n**) on which the twist-bend model is based. Subsequent variants, aiming at salvaging the initial association of the $N_X$ phase with the $N_{TB}$ model, introduced a polar aspect into the latter. [23] However, for this to be compatible with the fundamental hypothesis of the twist-bend model (i.e. the presence of an apolar nematic director **n**, whose elastic deformations are described exclusively in terms of the Frank-Oseen bend, splay and twist), the polarity has to have negligible effects on the full rotational symmetry about **n** and also negligible influence on the elastic properties of the medium. Such "phantom polarity", $\mathbf{P}_{ph}$, is then necessarily defined in terms of the bend vector $\mathbf{P}_{ph} \sim \mathbf{B} = \mathbf{n} \times \nabla \times \mathbf{n}$ of the spatial modulation of the nematic director **n** and is therefore transverse to **n** which is twisting and bending. In other words, in the case of a $N_{TB}$ the polarity is a result of the deformation and not of the local molecular ordering! (see figure 1). More recent attempts, [24] with the same aim to reinstate $N_{TB} = N_X$, include direct stipulation of polar molecular ordering and the recognition of **m** as the only local symmetry axis, and therefore the only uniquely defined director of the phase. In summary, such revisions of the $N_{TB}$ model essentially adopt the defining elements of the $N_{PT}$ model[22] albeit in a physically incoherent and self-contradictory way, while keeping the name of $N_{TB}$, thus casting doubt on and generating confusion about Meyer's elegant and physically clear original proposal. Specifically, if a nematic director **n** does not exist in the $N_{TB}$, as a result of the presence of a polar director **m** normal to the helix axis, then what is it that twists and bends? As a result of such continuing revisions of the $N_{TB}$ model to account for new experimental data in the $N_X$ phase, the original notion of the $N_{TB}$ is becoming obscure despite the wealth of experimental knowledge accumulated in the last decade, extending well beyond the original CB-n-CB dimers [25,26]. In summary the reluctance to accept the molecular organization in the $N_{PT}$ phase as a valid description of the $N_X$ phase has prevented a clear picture of this new nematic state from emerging. For example, from the most recent literature[27]: "...the key parameters that define the heliconical structure of the in $N_{TB}$ phase at the nanometer scale are not yet clearly understood. Thus, the complexity of the heliconical $N_{TB}$ structure must be simplified or minimized to deeply analyze the inherent structural properties of the phase." However, Figure 1, in conjunction with Meyer's 1973 explication, clearly defines the molecular organization in the $N_{TB}$ on the nanometer scale—it is essentially (aside from the "phantom polarity" described above) that of a uniaxial, apolar nematic. Similarly Vanakaras and Photinos [21,22] have described the molecular organization on the nanometer scale of the $N_{PT}$, i.e., the organization that applies in the $N_X$, the lower temperature nematic phase of the odd CB-Cn-CB dimers (Figure 2).

Microscopically, the difference between the $N_{TB}$ and the $N_{PT}$ models is reflected directly by the local molecular ordering (respectively, locally uniaxial and apolar, $\mathbf{n} \leftrightarrow -\mathbf{n}$, with twist-bend deformation of the nematic director and bend-associated phantom polarity, vs locally polar with a purely twisting polar director) and involves differences in the order of magnitude of the spatial modulation (pitch) and the polarity normal to the helix axis. The situation bares some similarity to the difference between cholesterics and nematics, as described by de Gennes[28] on replacing in his example, the long pitch cholesterics (twisted nematics) by the $N_{TB}$ and the short pitch cholesterics by the $N_{PT}$:

> "A perhaps more rigorous way to think of the difference between cholesterics and nematics is to use a comparison with phase transitions: when subjected to a small external magnetic field a paramagnetic phase acquires a small but non-zero macroscopic magnetization. It has the same symmetry as a ferromagnetic phase <u>although it is locally still very close to the initial paramagnetic state we started with</u>. If one increases the field enough, and in well chosen conditions, we know that we can drive the system continuously to a state that is truly ferromagnetic. Whether the paramagnetic phase subjected to a magnetic field should be considered as ferromagnetic or not is purely a matter of order of magnitude. Similarly, chirality acts as a field on the natural twisting tendency. The natural twist being almost always small on a molecular scale ($q_0 \sim 10^{-2}, 10^{-3}$), we are in the small field limit (i.e. the idea of cholesterics as twisted nematics is basically correct). On the other hand, the cholesteric state is really an original state of matter, and short-pitch cholesterics have probably little to do with nematics (i.e. the equivalent of the ferromagnetic phase). In fact, as we shall see, cholesterics could well be classified with smectics."

Accordingly, the idea of the $N_X$ phase as a twist-bend nematic would be basically correct if the modulation were in the macroscopic regime. The fact that it is not, but rather on the molecular length scale, points to a different state of matter (with substantial polar ordering, no nematic director, etc, as proposed in the $N_{PT}$ model) and even calls into question its very classification as a nematic [25,29]—the so-called "fifth type of nematic." [14].

There are clear and measurable differences in the molecular physics implied by the $N_{PT}$ and $N_{TB}$ models. In the $N_{PT}$ phase there is transverse polar molecular order, the twisting entity is the polar director **m**, the driving force of twist is entropic (the polar molecular packing of the V-shaped dimers), and the local phase symmetry is $C_2$. In contrast, the polarity in the $N_{TB}$ phase is necessarily negligible ("phantom"), with the polar molecular correlations originating from the flexopolarization coupling to the bend deformation, the twisting and bending entity is the nematic director **n**, the driving mechanism of twist and bend is based on deformation elasticity, as described by the Frank-Oseen formulation, extended with flexopolarization, and the local symmetry is essentially that of a uniaxial apolar nematic, so that a nematic director **n** can be defined. Furthermore, the $N_{TB}$ model implies a common value of the "tilt" angle for the principal axes of all molecular tensors whereas in the $N_{PT}$ model the "tilt" angle for the principal axes of molecular tensors is segment-dependent and differs from one tensor property to another. Finally, a physical property that is measurable by NMR is the doubling of spectral lines associated with prochiral sites (enantiotopic discrimination); this serves as a definitive signature of the $N_X$ phase and has been used to clearly identify the N-$N_X$ phase transition [12,13,20,30]. As shown in[20], the $N_{PT}$ model accounts for enantiotopic discrimination in small prochiral solutes whereas the $N_{TB}$ model does not. Extension to flexible and more extended solutes reveals further information on the molecular ordering in the $N_X$ phase and these new observations reinforce the correct description of the $N_X$ phase as the $N_{PT}$ model as shown in ref. [31]. Therein, arguments are also presented regarding the identification of the high temperature nematic phase (N). The N phase of the odd members of the CB-n-CB dimers is not a common (uniaxial apolar) nematic phase but rather a phase formed by molecular aggregates/clusters having the same structure as the $N_X$ domains albeit of much smaller spatial extent.

Returning to R.B. Meyer's original proposal[1] in 1973 of the NTB, we would like to conclude with three points:
- Meyer's proposal represents a pioneering achievement because it demonstrated theoretically for the first time that form-chirality can exist in simple nematics composed of <u>achiral</u> molecules. The hypothesis of having chiral self-organization in LC phases of achiral molecules was verified experimentally two decades later in various phases formed by bent-core molecules. [32]
- The clarity of Meyer's original description, as well as several elaborations thereof [33,34] founded on the Frank-Oseen elasticity theory of nematics, make a sharp contrast with subsequent obscure modifications to his concept, apparently aimed at salvaging the unfortunate identification of the $N_x$ with the $N_{TB}$.
- The $N_{TB}$ phase that Meyer so elegantly predicted may be identified experimentally in the future, as has often been the case with LC phases allowed by symmetry and that were eventually found experimentally. However, the search for the true $N_{TB}$ phase is not facilitated by the misuse of its name for other phases and/or the warping of its features to fit experiment.

Lastly, paraphrasing the quotation from Meyer's 1973 notes at the beginning of this article: No helical twist-bend structure has been demonstrated in chiral or normal nematics as of this writing.